\documentclass{ws-procs9x6}
\usepackage{graphicx,color}
\usepackage{epsfig,multirow}

\begin{document}

\title{Initial temperature of the strongly interacting Quark Gluon Plasma created at RHIC}

\author{M. CSAN\'AD}

\address{Department of Atomic Physics, E\"otv\"os University,\\
Budapest, P\'azm\'any P. s. 1/a, H-1117, Hungary\\
$^*$E-mail: csanad@elte.hu}

\begin{abstract}
A 1+3 dimensional solution of relativistic hydrodynamics is analyzed in this paper.
Momentum distribution and other observables are calculated from the solution and
compared to hadronic measurements from the Relativistic Heavy Ion Collider (RHIC). The
solution is compatible with the data, but only the freeze-out point of the evolution is determined.
Many equation of states and initial states (initial temperatures) are valid with the same freeze-out distribution, thus
the same hadronic observables. The observable that would distinguish between these initial temperatures is
momentum distribution of photons, as photons are created throughout the evolution of the fireball created in
RHIC collisions. The PHENIX experiment at RHIC measures such data via low invariant mass $e^+e^-$ pairs.
Average temperature from this data is $T=221\pm23$(stat)$\pm18$(sys) MeV, while a model calculation with initial temperature
$T_{init}$ = 370 MeV agree with the data.
\end{abstract}


\bodymatter

\section{Perfect fluid hydrodynamics}
In the last several years it has been revealed that the strongly interacting Quark Gluon Plasma produced~\cite{Adcox:2004mh} in the collisions of the
Relativistic Heavy Ion Collider (RHIC) is a nearly perfect fluid~\cite{Lacey:2006bc}, i.e.\ it can be described with perfect fluid hydrodynamics.

Perfect fluid hydrodynamics is based on local conservation of entropy or number density ($n$), energy-momentum density ($T^{\mu\nu}$).
The fluid is perfect if the energy-momentum tensor is diagonal in the local rest frame, i.e.\ 
viscosity and heat conduction are negligible. This can be assured if $T^{\mu\nu}$ is chosen as $T^{\mu\nu}=(\epsilon+p)u^\mu u^\nu-pg^{\mu\nu}$,
where $u^\mu$ is the flow field in the fluid, $\epsilon$ is energy density, $p$ is pressure and $g^{\mu\nu}$ is the metric tensor, diag(1,-1,-1,-1).
The conservation equations are closed by the equation of state, which gives the relationship between $\epsilon$ and $p$. Typically $\epsilon = \kappa p$ is chosen,
where the proportionality ``constant'' $\kappa$ may depend on temperature $T$, which in turn is connected to the density $n$ and pressure $p$ via $p=nT$.

The exact, analytic result for hydrodynamic solutions is, that the hadronic observables do not depend on the initial state or the dynamical equations separately,
just through the final state~\cite{Csanad:2009sk,Csanad:2009wc}. Thus if we fix the final state from the data, the equation of state can be anything that is
compatible with the particular solution. This is the framework of several hydro solutions as detailed in the next paragraph.

Many solve the above equations numerically, but there are only a few exact solutions. Historically the first is the implicit 1+1 dimensional accelerating solution
of Landau and Khalatnikov~\cite{Landau:1953gs,Khalatnikov:1954aa,Belenkij:1956cd}. Another renowned 1+1 dimensional solution of relativistic hydrodynamics was found by
Hwa and Bjorken\cite{Hwa:1974gn,Chiu:1975hw,Bjorken:1982qr}: it is simple, explicit and exact, but accelerationless.

Important are solutions~\cite{Csorgo:2006ax,Bialas:2007iu} which are explicit and describe a relativistic acceleration, i.e. combine the
properties of the Landau-Khalatnikow and the Hwa-Bjorken solutions. With these one can have an advanced estimate on the energy density~\cite{Csanad:2007iv},
but investigation of transverse dynamics is not possible by these solutions.

The only exact 1+3 dimensional relativistic solution, from which observables like momentum distribution, correlation function and elliptic flow were
 calculated~\cite{Csanad:2009wc} is the one in ref.~\cite{Csorgo:2003ry}. Observables from this solution were computed and compared
to data in ref.~\cite{Csanad:2009wc}. 

\section{The analyzed solution}
The analyzed solution~\cite{Csorgo:2003ry} describes an ellipsoidally symmetric expansion.
The ellipsoids are given by constant values of the scale variable $s$:
\begin{align}
s=\frac{r_x^2}{X(t)^2}+\frac{r_y^2}{Y(t)^2}+\frac{r_z^2}{Z(t)^2},
\end{align}
here $X(t)$, $Y(t)$, and $Z(t)$ are time dependent scale parameters (axes of the $s=1$ ellipsoid),
only depending on the time $t$. Spatial coordinates are $r_x$, $r_y$, and $r_z$.
The velocity-field is described by a Hubble-type expansion:
\begin{align}
u^\mu (x) = \gamma \left(1, \frac{\dot X(t)}{X(t)}r_x, \frac{\dot Y(t)}{Y(t)}r_y, \frac{\dot Z(t)}{Z(t)}r_z\right),
\end{align}
where $x$ means the four-vector $(t,r_x,r_y,r_z)$, and $\dot X(t) = dX(t)/dt$, similarly for $Y$ and $Z$.
The $\dot X(t)=\dot X_0$, $\dot Y(t)=\dot Y_0$, $\dot Z(t)=\dot Z_0$ (i.e. all are constant) criteria must
be fulfilled, ie.\ the solution is accelerationless. This is one of the drawbacks of this solution.

\begin{figure}
	\centering
		\includegraphics[width=0.90\textwidth]{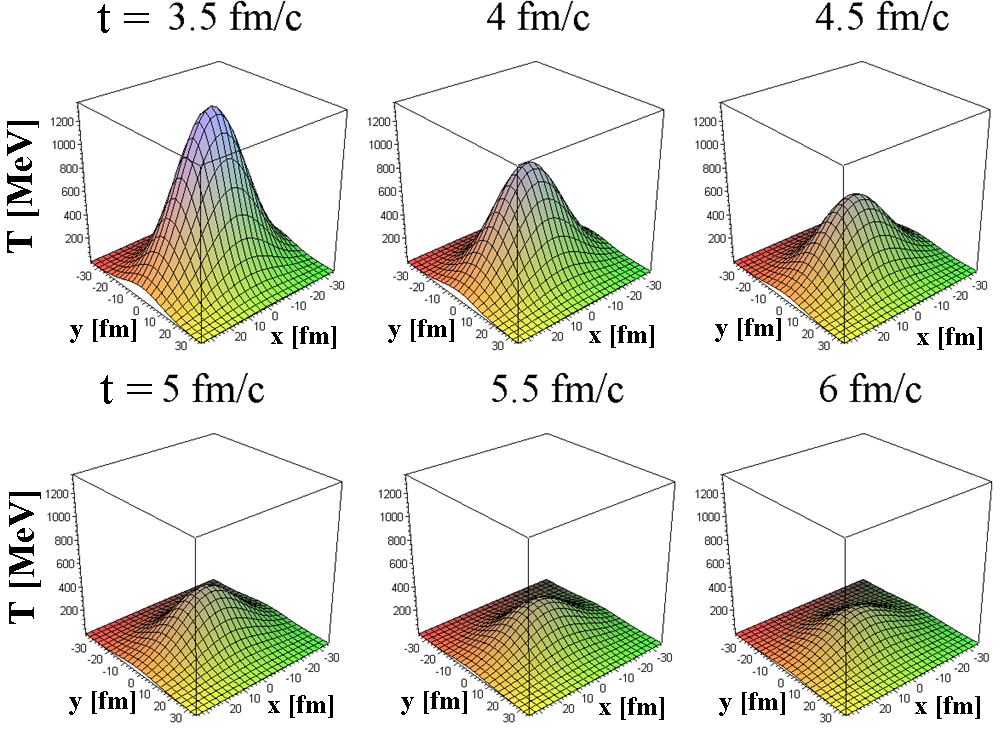}
	\caption{Temperature distribution in the transverse plane ($x$-$y$) is shown for various times with an example parameter set. The fireball is the hottest in the center
	and it cools down as time elapses.}
	\label{f:tempdist}
\end{figure}

The temperature $T(x)$ and number density $n(x)$ are:
\begin{align}
n(x)&=n_0\left(\frac{\tau_0}{\tau}\right)^3 \nu(s), \\
T(x)&=T_0\left(\frac{\tau_0}{\tau}\right)^{3/\kappa} \frac{1}{\nu(s)},\label{e:temp}\\
p(x)&=p_0\left(\frac{\tau_0}{\tau}\right)^{3(\kappa+1)/\kappa},
\end{align}
where $\tau$ is the proper time, $s$ is the above scaling variable, $\nu(s)$ is an arbitrary function, while $n_0=n|_{s=0,\tau=\tau_0}$, $T_0=T|_{s=0,\tau=\tau_0}$
and $p_0=p|_{s=0,\tau=\tau_0}$ with $p_0 = n_0 T_0$ (hence $p$ does not depend on the spatial coordinates only $\tau$). Furthermore, $\tau_0$ is the time of the freeze-out,
thus $T_0$ is the central freeze-out temperature. The parameter $\kappa$ is arbitrary, i.e. any value of $\kappa$ yields a solution.
The function $\nu(s)$ is chosen as:
\begin{align}
\nu(s)=e^{-bs/2},
\end{align}
where $b$ is then the temperature gradient. If the fireball is the hottest in the center, then $b<0$. An example time evolution of the temperature distribution is shown in
fig.~\ref{f:tempdist}

\section{Hadronic observables}
The picture widely used in hydro models is that the pre freeze-out (FO) medium is described by hydrodynamics, and the post FO medium is that
of observed hadrons. In our framework we assume that the freeze-out can happen at any proper time, e.g in case of a self-quenching effect or
if the phase space evolution is that of a collisionless gas. See details in ref.~\cite{Csanad:2009wc}. The hadronic observables can be extracted
from the solution via the phase-space distribution at the FO. This will correspond to the hadronic final state or source distribution $S(x,p)$.

We do not need to fix a special equation of state, because the same final state can be achieved with different equations of state or initial
conditions~\cite{Csanad:2009sk}. Thus the hadronic observables do not restrict the value of $\kappa$.

The first calculated observable is invariant transverse momentum distribution $N_1(p_t)$ of a particle with mass $m$~\cite{Csanad:2009wc}:
\begin{align}
\label{e:n1pt}
N_1(p_t)= &2\pi\overline{N}\;\overline{V} \left(m_t-\frac{p_t^2(T_{\rm eff}-T_0)}{m_tT_{\rm eff}}\right)
          \exp\left[-\frac{m_t^2+m^2}{2m_tT_0}-\frac{p_t^2}{2m_tT_{\rm eff}}\right],\nonumber
\end{align}
with the following auxiliary quantities:
\begin{align}
\overline{N}&=\mathcal{N} n_0 \left(\frac{2T_0\tau_0^2\pi}{m_t} \right)^{3/2},\\
\overline{V}&=\sqrt{\left(1-\frac{T_0}{T_x}\right)\left(1-\frac{T_0}{T_y}\right)\left(1-\frac{T_0}{T_z}\right)},\\
\frac{1}{T_{\rm eff}}&=\frac{1}{2}\left(\frac{1}{T_x}+\frac{1}{T_y}\right).
\end{align}
Furthermore, $T_x$ ,$T_y$, $T_z$ are the effective temperatures, i.e. inverse logarithmic slopes of the distribution:
\begin{align}
T_x&=T_0+\frac{m_t T_0\dot X_0^2}{b(T_0-m_t)}, \\
T_y&=T_0+\frac{m_t T_0\dot Y_0^2}{b(T_0-m_t)}, \\
T_z&=T_0+\frac{m_t T_0\dot Z_0^2}{b(T_0-m_t)},
\end{align}
where $\dot X_0$, $\dot Y_0$ and $\dot Z_0$ are the (constant) expansion rates of the fireball, $T_0$ its central temperature
at FO and $b$ the temperature gradient.

We also calculate the elliptic flow, which describes the azimuthal asymmetry of the momentum distribution~\cite{Csanad:2009wc}:
\begin{align}
\label{e:v2}
v_2(p_t)=\frac{I_1(w)}{I_0(w)}
\end{align}
where $I_0$, and $I_1$ are the modified Bessel functions while
\begin{align}
w=\frac{p_t^2}{4m_t}\left(\frac{1}{T_y}-\frac{1}{T_x}\right).
\end{align}
See details of the calculation in ref.~\cite{Csanad:2009wc}. The formula for $v_2$ gives back previously found formulas of non-relativistic solutions~\cite{Csorgo:2001xm}
and relativistic solutions~\cite{Csanad:2003qa,Csanad:2005gv}. Also the formula for $N_1(p_t)$ is similar to results of the previously mentioned papers.

Third observable we calculate is the two-particle Bose-Einstein (HBT) correlation radii of identical bosons~\cite{Weiner:2000wj}:
\begin{align}
R_x^2&=\frac{T_0\tau_0^2(T_x-T_0)}{M_tT_x},\\
R_y^2&=\frac{T_0\tau_0^2(T_y-T_0)}{M_tT_y},\\
R_z^2&=\frac{T_0\tau_0^2(T_z-T_0)}{M_tT_z},
\end{align}
where $M_t$ is the transverse mass belonging to the average momentum $K=0.5(p_1+p_2)$ of the pair, which is (at mid-rapidity) $M_t=0.5\left(m_{t,1}+m_{t,2}\,\right)$. The
$m_{t,1}$, and $m_{t,2}$ quantities are the transverse masses, the $T_{x}$, $T_{y}$, and $T_{z}$ are the effective temperatures belonging to the average momentum (i.e.\ 
here $T_x=T_x|_{M_t}$). The calculations are detailed in ref~\cite{Csanad:2009wc}. To compare the HBT radii with the data the Bertsch-Pratt~\cite{Pratt:1986cc} frame is to
 be used. It has three axes: the \emph{out} is the direction of the average transverse momentum of the pair, the \emph{long} direction is equal to the direction \emph{z},
 and the \emph{side} direction is orthogonal to both of them. The result for $R_{\rm out}$, $R_{\rm side}$ and $R_{\rm long}$ is:
\begin{align}
R_{\rm out}^2=R_{\rm side}^2&=\frac{R_x^2+R_y^2}{2}, \label{e:ros}\\
R_{\rm long}^2&=R_z^2.
\end{align}
Clearly in this solution the out and side radii are equal. This can be attributed to the instantaneous freeze-out; a non-zero freeze-out duration would make
$R_{\rm out}^2$ bigger by a term of $\Delta \tau^2 p_t^2/E^2$. Supported by the data, we use the $\Delta \tau = 0$ approximation in our solution,
which corresponds to instantaneous freeze-out.

An important consequence of the above results is that neither spectra nor elliptic flow nor correlation radii depend on the EoS itself, only through the final state
parameters. If we determine for example $T_0$, the freeze-out central (at the center means here $r_x=r_y=r_z=0$) temperature,
$\kappa$ or the initial temperature $T_{\rm initial}$ still cannot be calculated. We only know that they are connected through the consistency condition
$T_{\rm initial} = T_0 (\tau_0/\tau_{\rm initial})^{3/\kappa}$, see eq.~(\ref{e:temp}), i.e. they can be co-varied (softer EoS requires smaller initial
temperature for a given freeze-out proper-time). Thus $\kappa$ or $T_{\rm initial}$ has to be determined from another measurement, e.g. the spectrum of thermal photons.

\section{Comparing the hadronic observables to RHIC data}
The above results were compared in ref.~\cite{Csanad:2009wc} to PHENIX data of 200 GeV Au+Au collisions. Above formulas were fitted to describe spectra and HBT positive pion
data~\cite{Adler:2003cb,Adler:2004rq} (0-30\% centrality) and elliptic flow data~\cite{Adler:2003kt} for $\pi^\pm$, $K^\pm$, p and
$\overline{\rm p}$ particles (0-92\% centrality).

\begin{figure}
	\centering
		\includegraphics[width=0.48\textwidth]{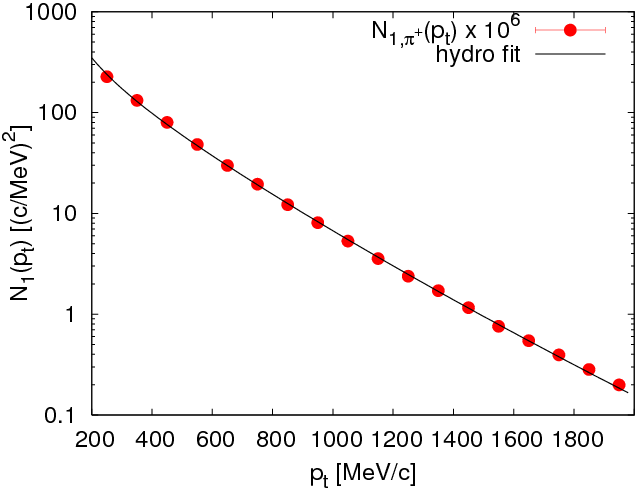}
		\includegraphics[width=0.48\textwidth]{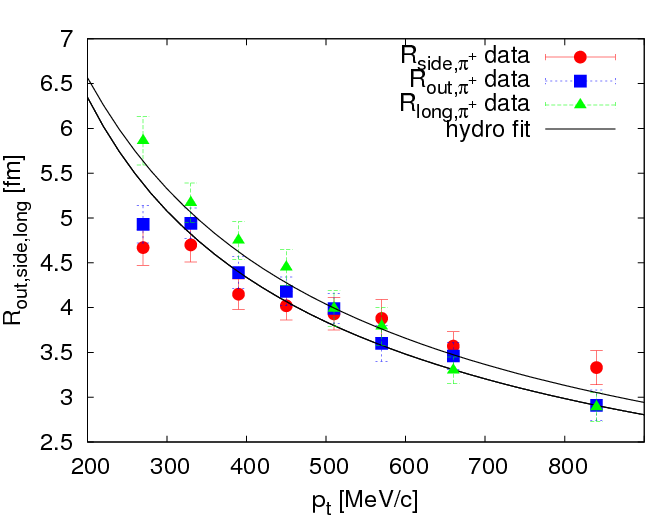}
		\includegraphics[width=0.48\textwidth]{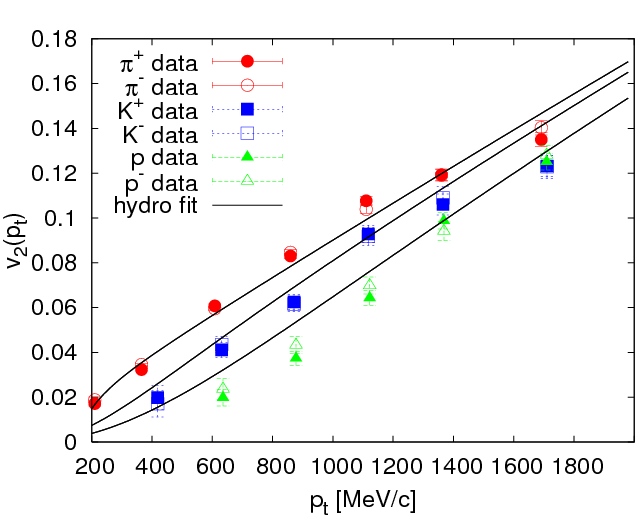}
	\caption{Fits to 0-30\% centrality PHENIX Au+Au spectra~\cite{Adler:2003cb} (top left) HBT radii~\cite{Adler:2003kt} (top right)
	and 0-92\% centrality PHENIX Au+Au elliptic flow~\cite{Adler:2004rq} (bottom). See details of the fit in ref.~\cite{Csanad:2009wc}.}
	\label{f:fits}
\end{figure}

The fit results are shown in fig.~\ref{f:fits}, see details of the fit in ref.~\cite{Csanad:2009wc}. Important are the following: central freeze-out temperature
$T_0$ is around 200 MeV for both datasets (with an error of 7 MeV), and the fireball is colder away from the center. The expansion eccentricity is positive, it tells us that
the expansion is faster in-plane. Because of the Hubble-flow this means that the source is in-plane elongated, similarly to the result of ref.~\cite{Csanad:2008af}.
The freeze-out happens at a proper-time of $\tau_0 = 7.7\pm 0.8$ fm$/c$.

\begin{figure}
	\centering
	  \includegraphics[width =0.80\textwidth]{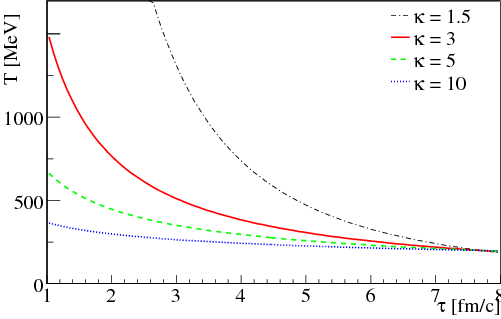}
	\caption{	Time dependence of the central temperature of the fireball, from eq.~(\ref{e:temp}) is shown for different $\kappa$ values.
	In reality $\kappa$ may change with time, we show here the curves only for fixed $\kappa$ values. Assuming an
	average of $\kappa=10$~\cite{Lacey:2006pn} one gets an initial temperature of 370 MeV at $t_{\rm initial}$=1 fm$/c$, in agreement
	with PHENIX measurements~\cite{Adare:2008fqa}.}
	\label{f:temp}
\end{figure}

Our fit parameters describe the fireball at the freeze-out. However, the solution is time-dependent, most importantly the temperature
depends on time as described by eq.~(\ref{e:temp}). We plotted the time-dependence of the central temperature in fig.~\ref{f:temp} for several
values of $\kappa$, i.e. several EoS'. From this, assuming for example an average $\kappa$ of 10, see ref.~\cite{Lacey:2006pn}, one can also calculate
the initial central temperature of the fireball based on eq.~(\ref{e:temp}):
\begin{align}
T_{\rm initial} = T_0 \left(\frac{\tau_0}{\tau_{\rm initial}}\right)^{3/\kappa}
\end{align}
This yields 370 MeV at $t_{\rm initial}$=1 fm$/c$ (note that $t=\tau$ at the center).

One still would like to determine the value for $\kappa$ and the initial temperature from experimental observables. The key are thermal photon invariant momentum
distributions, because photons are not suppressed by the medium, they are produced according to the local temperature at all times. If one could measure the yield of
thermal (direct) photons, one could compare it to time-integrated hydrodynamic results and determine EoS and initial temperature. In the next section thus experimental
results on direct photons are reviewed.

\section{Direct photons at the PHENIX experiment at RHIC}
Thermal photons from the partonic phase are predicted to be the dominant source of direct photons for $1<p_t<3$ GeV/$c$ in Au+Au collisions
at the Relativistic Heavy Ion Collider (RHIC)~\cite{Turbide:2003si}. The measurement of direct photons in this kinematic domain is however very difficult due to the
background from hadronic decay photons. At PHENIX, an alternative approach is used~\cite{Adare:2008fqa}. The idea is that any source of high energy photons can also
emit virtual photons which then convert to $e^+e^-$ pairs.

The relation between photon production and the associated $e^+e^-$ pair production is then~\cite{Adare:2008fqa,Adare:2009qk}
\begin{align}
\frac{d^2n_{ee}}{dm} = \frac{2\alpha}{3\pi m} \sqrt{1-\frac{4m_e^2}{m^2}}\left( 1+\frac{2m_e^2}{m^2} \right) S  dn_{\gamma} \label{e:conv}
\end{align}
Here $\alpha$ is the fine structure constant, $m_e$ and $m$ are the masses of the electron and the $e^+e^-$ pair respectively,
and $S$ is a process dependent factor.

\begin{figure}
	\centering
  \includegraphics[width=0.85\linewidth]{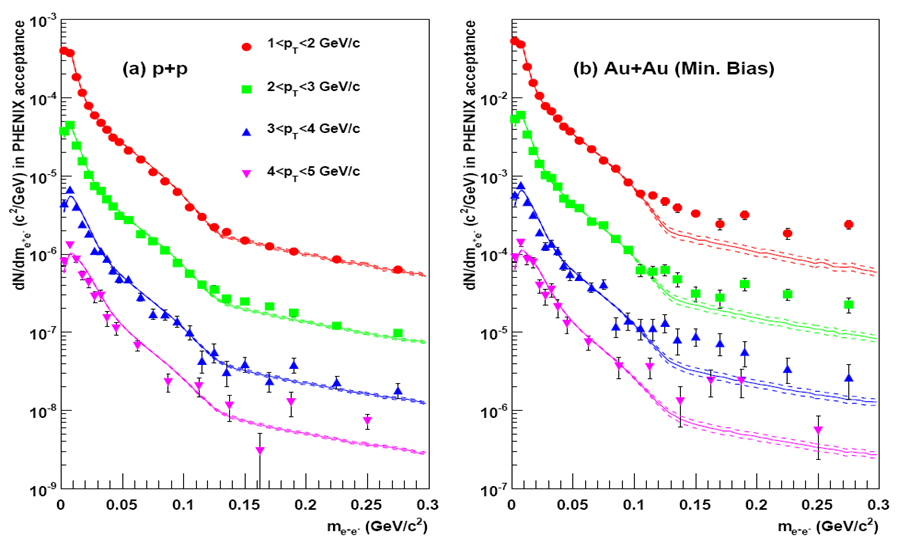}
  \includegraphics[width=0.85\linewidth]{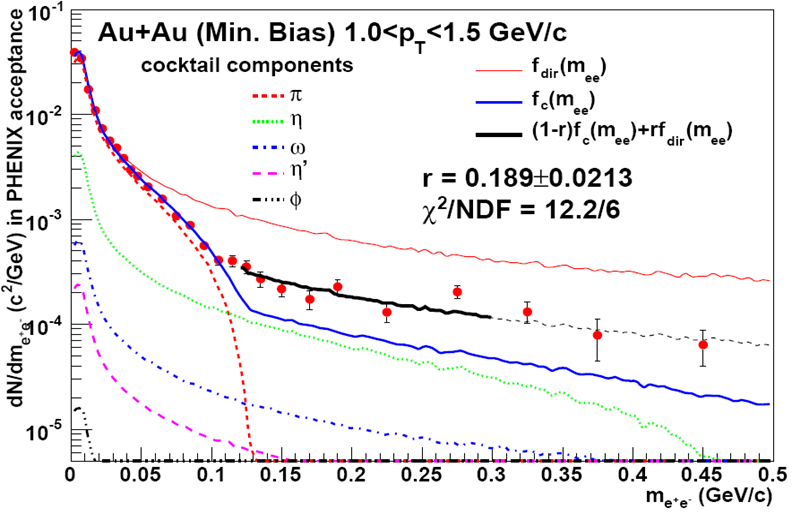}
  \caption{The $e^+e^-$ pair invariant mass distributions in $p+p$ and minimum bias Au+Au collisions from ref.~\cite{Adare:2008fqa} is shown in the top panel. Mass
  distribution of electron pairs measured in Au+Au minimum bias events~\cite{Adare:2008fqa} is shown for $1.0<p_t<1.5$ GeV/$c$ in the bottom panel.
  The fit is explained in the text and originally in ref.~\cite{Adare:2008fqa}.}\label{f:twocompfit}
\end{figure}

Invariant mass distribution of $e^+e^-$ pairs was measured thus in PHENIX~\cite{Adare:2008fqa,Adare:2009qk} for $m_{ee} < 300$ MeV/$c^2$ and for $1< p_z< 5$ GeV/$c$
in Au+Au and $p+p$ collisions at $\sqrt{s_{NN}}=200$ GeV. Top panel of fig.~\ref{f:twocompfit} shows the measured~\cite{Adare:2008fqa} mass spectra of $e^+e^-$ pairs in
$p+p$ and Au+Au collisions for different ranges of $p_t$, comparing them to expected yields from dielectron decays of hadrons, calculated using a Monte Carlo hadron decay
generator based on meson production as measured by PHENIX~\cite{Adare:2008asa}. The Au+Au data show a relatively high excess above the hadronic background, which indicates 
an the production of virtual photons in Au+Au collisions. PHENIX assumed that the excess is entirely due to internal conversion of direct photons and deduce the real direct 
photon yield from the $e^+e^-$ pair yield using eq.~(\ref{e:conv})~\cite{Adare:2008asa}.

In order to calculate direct photon yields the $f(m) = (1-r)f_{c}(m) + r f_{dir}(m)$ fit function was used to the mass distribution in each $p_t$ bin separately,
where $f_{c}(m)$ is the mass distribution from the hadronic decays (shown in the top panel of fig.~\ref{f:twocompfit} with solid lines), and $f_{dir}(m)$ is the expected
shape of the direct photon internal conversion, and $r$ is the fit parameter. Such a fit is shown in the bottom panel of fig.~\ref{f:twocompfit} for one $p_t$ bin. The
direct photon yield can be calculated then using the direct photon fraction $r$ as $dN^{\rm direct}(p_t)= r \times dN^{\rm inclusive}(p_t)$ where $dN^{\rm inclusive}(p_t)$
is the yield of all photons (direct and decay inclusive). The inclusive photon yield can be calculated as
$dN_\gamma^{\rm inclusive}=N_{ee}^{\rm data} \times (dN_{\gamma}^{\rm cocktail}/N_{ee}^{\rm cocktail})$, where $N_{ee}^{\rm data}$ and
$N_{ee}^{\rm cocktail}$ are the measured and cocktail $e^+e^-$ pair yields and $dN_{\gamma}^{\rm cocktail}$ is the yield of photons from the cocktail.

\begin{figure}
	\centering
  \includegraphics[width=0.85\linewidth]{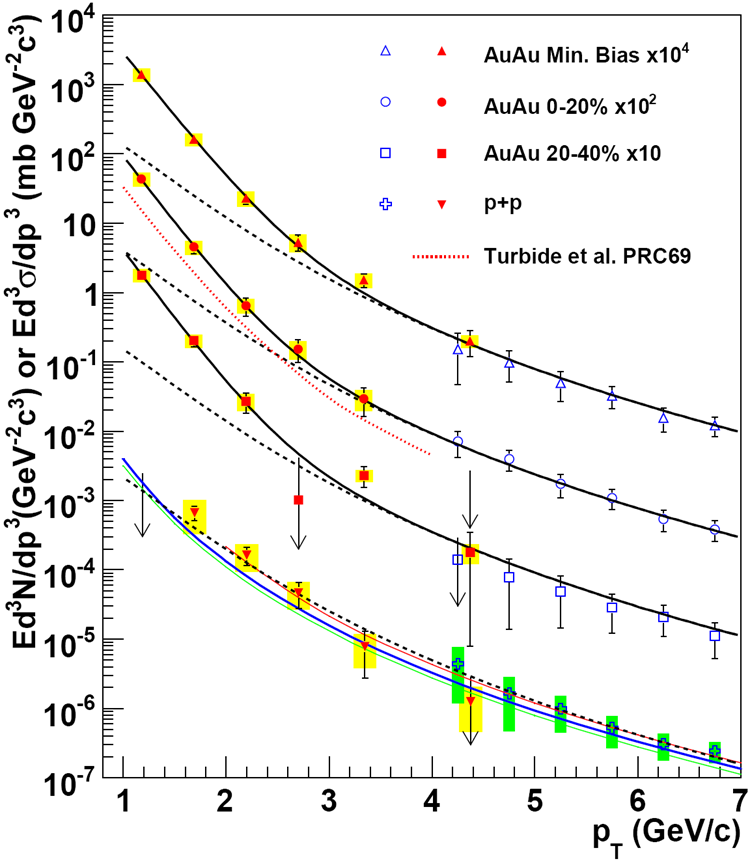}
  \caption{Invariant cross section ($p+p$) and invariant yield (Au+Au) of direct photons is shown, taken from ref.~\cite{Adare:2008fqa}.
  The red points are from the analysis presented in ref.~\cite{Adare:2008fqa} and blue points are from~\cite{Adler:2005ig,Adler:2006yt}. 
  Dashed black curves are modified power-law fits to the $p+p$ data~\cite{Adare:2008fqa}, while the solid black curves are exponential plus modified power-law
  fits~\cite{Adare:2008fqa}. The red dashed curve is the direct photon spectrum in central Au+Au collisions in ref~\cite{Turbide:2003si}.}\label{f:photonyield}
\end{figure}

The obtained direct photon spectra~\cite{Adare:2008fqa} are shown in fig.~\ref{f:photonyield} and compared to similar data of refs.~\cite{Adler:2005ig,Adler:2006yt}.
The direct photon yields are consistent with a NLO pQCD calculation in $p+p$.
The shape of the direct photon spectra above a binary collisions scaled $p+p$ spectrum is exponential in $p_t$, with an inverse slope $T=221\pm23$(stat)$\pm18$(sys) MeV
in central Au+Au collisions~\cite{Adare:2008fqa}. This temperature can be regarded as a time-average of the temperature of the fireball, represents thus an experimental
lower limit to the initial temperature. The shape of the thermal photon spectrum from a model calculation with initial temperature $T_{init}$ = 370 MeV agrees with the
data~\cite{Adare:2008fqa}.

\section{Summary}
Exact parametric solutions of perfect hydrodynamics were long searched for in order to describe the matter produced in heavy ion collisions at RHIC.
We extracted hadronic observables from the relativistic, 1+3 dimensional, ellipsoidally symmetric, exact solution of ref.~\cite{Csorgo:2003ry}.
We calculated momentum distribution, elliptic flow and Bose-Einstein correlation radii from the solution. We compared the results to 200 GeV Au+Au
PHENIX data~\cite{Adler:2003cb,Adler:2003kt,Adler:2004rq}. The solution is compatible with the data. If using an experimentally determined average EoS of
$\kappa\approx10$~\cite{Lacey:2006pn}, our results yield approximately 370 MeV at $\tau_{\rm initial}$=1 fm$/c$, in agreement with recent PHENIX
photon measurements~\cite{Adare:2008fqa}. The time-average of the temperature in these collisions is $T=221\pm23$(stat)$\pm18$(sys) MeV~\cite{Adare:2008fqa},
which is a lower limit of the initial temperature. From detailed comparisons with hydrodynamical models of direct photon emission, PHENIX concluded that
$T_{init}$ = 300 MeV is the lowest possible initial temperature at 1 fm/c, that is consistent with such an average slope parameter. In the future we will
compare hadron and photon observables simultaneously to the analyzed analytic hydrodynamic solution.

\section*{Acknowledgments}
The author is thankful for the support of the organizers of the Gribov-80 Memorial Workshop, it was a pleasure to participate in a conference of so many well-known
scientists, in such a pleasent and friendly atmosphere. The author also gratefully acknowledges the support of the Hungarian OTKA grant NK 73143.

\bibliographystyle{ws-procs9x6}
\bibliography{../../../master}

\end{document}